\documentclass[conference]{IEEEtran}
\IEEEoverridecommandlockouts
\usepackage{cite}

\ifCLASSINFOpdf
   \usepackage[pdftex]{graphicx}
   \DeclareGraphicsExtensions{.pdf,.jpeg,.png}
\else
  \usepackage[dvips]{graphicx}
  \DeclareGraphicsExtensions{.eps}
\fi

\usepackage{textcomp}
\usepackage{xcolor}

\usepackage{epstopdf}
\usepackage{amsthm}
\usepackage{lipsum}
\usepackage{color}
\usepackage{algorithm}
\usepackage{algorithmic}
\usepackage{enumerate}

\usepackage{etoolbox}  
\makeatletter
\patchcmd{\algorithmic}{\addtolength{\ALC@tlm}{\leftmargin} }{\addtolength{\ALC@tlm}{\leftmargin}}{}{}
\usepackage{amssymb}

\ifCLASSINFOpdf
   \usepackage[pdftex]{graphicx}
\else
\fi
\usepackage{cite}
\usepackage[cmex10]{amsmath}
\usepackage{algorithmic}
\usepackage{array}
\usepackage{url}
\usepackage{makeidx}
\usepackage{verbatim}
\usepackage{subfigure}

\usepackage{multirow}
\newcommand{\beqn}{\begin{eqnarray}}
\newcommand{\eeqn}{\end{eqnarray}}
\newcommand{\bse}{\begin{subequations}}
\newcommand{\ese}{\end{subequations}}

\begin{document}

\title{A Quantum Neural Network Regression for Modeling Lithium-ion Battery Capacity Degradation

\author{Anh Phuong Ngo$^1$, Nhat Le$^1$, Hieu T. Nguyen$^1$, Abdullah Eroglu$^1$, and Duong T. Nguyen$^2$\\ 
$^1$\textit{Dept. of Electrical \& Computer Eng., North Carolina A\&T State University, Greensboro, NC 27411, USA} \\
$^2$\textit{Dept. of Electrical \& Computer Eng, Arizona State University, Tempe, AZ 85281, USA}\\
Email: \textit{\{ango1, nmle\}@aggies.ncat.edu, \{htnguyen1, aeroglu\}@ncat.edu, duongnt@asu.edu}
}
}

\maketitle
\begin{abstract}
Given  the high power density, low discharge rate, and decreasing cost, rechargeable lithium-ion batteries (LiBs) have found a wide range of applications  such as  power grid-level storage systems, electric vehicles, and mobile devices. 
Developing a framework to accurately model the nonlinear degradation process of LiBs, which is indeed a supervised learning problem, becomes an important research topic.
This paper presents a classical-quantum hybrid machine learning approach to capture the LiB degradation model that assesses battery cell life loss from operating profiles. 
Our work is motivated by recent advances in quantum computers as well as the similarity between neural networks and quantum circuits.
Similar to adjusting weight parameters in conventional neural networks, the parameters of the quantum circuit, namely the qubits' degree of freedom, can be tuned to learn a nonlinear function in a supervised learning fashion.  
As a proof of concept paper, our obtained numerical results with the battery dataset provided by NASA demonstrate the ability of the quantum neural networks in modeling the nonlinear relationship between the degraded capacity and the operating cycles. 
We also discuss the potential advantage of the quantum approach compared to conventional neural networks in classical computers in dealing with massive data, especially in the context of future penetration of EVs and energy storage.


\end{abstract}



{ 
\footnotesize

}

\section{Introduction}
Lithium-ion batteries (LiBs) are the dominant player in the battery market for electric vehicles (EVs), energy storage, and mobile devices thanks to their high energy densities, low cost, and long life cycle.
However, LiBs pose a concern about their capacity degradation which has a negative impact on the safety and reliability of these applications.
As a result, the estimation of battery cycling capacity is essential for the utilization of battery management system operation, economical and safety aspects, and life cycle assessment. Fundamentally, the battery capacity degradation estimation approaches are classified into two groups: model-based and data-driven approaches \cite{improved_softplus_ANN}.

On the one hand, the model-based approaches adopt parametric electrochemical processes to investigate the relationship between the cell capacity and cyclic aging of a lithium-ion battery \cite{bat_cap_degradation2021}.
However, mathematical modeling and parametric identification of the chemical properties of the elements in a LiB are irreversible and highly complex \cite{laresgoiti2015modeling}.
Consequently, a comprehensive result is difficult to be obtained with the model-based approaches.
Furthermore, every model-based technique is specifically developed for a certain type of LiB.
Therefore, its extensibility to another composition other than its origin is very limited \cite{review_on_ML_methods}.
On the other hand, the data-driven approaches are non-parametric.
They are based on the extraction of empirical observations of battery operation, such as voltage, current, temperature, and capacity measurements.
Hence, they are scalable and extensible to adapt to the variations in batteries' size and type. 
Furthermore, the rapid development of data collection techniques and computational processors has facilitated the practical application of data-driven methods.
Numerous data-based analytical methods have been applied in practice for the prediction of LiBs degradation are support vector machine, relevance vector machine,  Gaussian process functional regressions, and neural networks \cite{Classical_Battery_ML}. 
Among data-driven methods, artificial neural network (ANN) is a powerful machine learning tool with the capability to handle big data involving complex nonlinear systems \cite{improved_softplus_ANN, goodfellow2016deep}. 

With demand for EVs and electronic devices on the rise, LiBs as one of the essential components required for this mobility and connectivity boom has been formulating a big challenge of managing a huge amount of data.
Meanwhile, classical machine learning models usually run into a performance bottleneck trained on such kind of heavy tasks. 
Recent advances in quantum computing have further enabled quantum machine learning (QML) models for potentially achieving promising performance \cite{hybrid_qnn,quantum_kernel_ibm2022}.
In other words, QML models use quantum computers to boost up the power of machine learning \cite{quantum_info}.
Quantum neural network (QNN) is a widely-used QML model that composes of parameterized learnable quantum circuits \cite{q_neural_design}. 
In a QNN, the classical features are encoded into quantum states by using angle encoding. 
Quantum states are then used as input for a QNN model composed by layers of the learnable quantum circuit \cite{quantum_circuit_learning, qcl2}.
The quantum circuit is constructed by a series of parameterized rotation gates along axes and the measurements of qubits are decoded into classical values in the output layer.
This work develops a QNN regression model to predict the deterioration of battery capacity with the purpose of assessing potential quantum advantage in such kind of learning tasks for batteries.

This paper is organized as follows.
Section II presents the basics of how a Lithium-ion battery works and its capacity degradation model.
We will explain carefully the similarities and differences between QNN and ANN models in Section III. 
The numerical results for the QNN model are given in the Section IV.
Finally, Section V concludes the paper and discusses possible future research.

\section{Battery capacity degradation}
\subsection{Basics of Lithium-ion Battery Model}
LiB is a type of energy storage technology constructed of an anode, cathode, separator, electrolyte, and two current collectors, one positive and other negatives. 
LiB utilizes insertion reactions for the cathode and anode with lithium ions as the charge carriers. 
The anode delivers lithium ions to the cathode, which initiates a flow of electrons between the two components.
In simple words, a LiB employs its $Li^+$ ions as a key component of its electrochemistry \cite{bat_cap_degradation2021}.
An example of this type of electrochemical reaction is Lithium Manganese Oxide spinel $LiMnO_2$  illustrated as follows \cite{reason4bat_cap_degradation2020}:
$$
	\begin{cases}
		LiMn\rightarrow Li_{1-x}MnO_2 + xLi^+ + xe^-, & \text{at electrode}^+\\
		mC + xLi^+ + xe^- \rightarrow Li_xC_m, & \text{at electrode}^- \label{example_chemical_reaction}
	\end{cases}
$$

\subsection{Li-ion Battery Capacity Degradation Behavior}
The capacity degradation phenomenon is constituted by various factors: (i) formation property of the solid-electrolyte interface (SEI) layer and (ii) irreversible absorption of $Li^+$ ion at the host material.
As such, the thickness of the SEI layer keeps increasing over time, because it passively absorbs $Li^+$ ion during the charging and discharging processes along the battery life span. 
A LiB will become dead when its capacity degradation drops to a failure threshold defined by its manufacturer \cite{bat_cap_degradation2021}.
While the degradation of LiB can be theoretically explained by the loss of lithium ions and other active materials, it is difficult to link such molecular-level degradation processes to the operational pattern of energy storage, particularly the charging and discharging cycles \cite{laresgoiti2015modeling}. 
Machine learning with empirical datasets becomes a promising approach for estimating the LiB capacity deterioration

\subsection{Regression models for the Li-ion Battery Capacity Degradation Estimation}
The problem of modeling the degraded battery capacity induced by operational cycles can be considered as a regression problem in the field of supervised learning. 
Let $Y = (y_1,y_2,\cdots, y_n)^\top \in \mathbb{R}^n$ is historical capacity measurements of a LiB over a cycle $X = (x_1,x_2,\cdots,x_n)^\top$.
We need to construct a nonlinear mapping function $y= f_\theta(x)$ that corresponds the operational cycles $x$ to the remained capacity $y$ where $\theta$ is the vector of parameters in the hypothetical function $f_\theta(x)$.
The problem boils into finding the optimal $\theta^*$ that minimizes the loss function between the estimated $\hat{y} = f_\theta(x)$ and the measured capacity $y$ of the battery.
Two popular loss functions are Root Mean Square Error (RMSE) and Mean Absolute Percentage Error (MAPE):
\beqn
        RMSE = \sqrt{\frac{1}{N} \sum_{i=1}^N \left( y_i - \hat{y}_i \right)^2} \label{rmse}\\
        MAPE(\%) =  \frac{1}{N} \sum_{i=1}^N \frac{\left| y_i - \hat{y}_i  \right|}{y_i} \times 100
\eeqn
where $y_i$ is the real measured capacity and $\hat{y}_i$ is the predicted capacity at cycle $i \in N$.

\begin{figure}[t!]
    \centering
	\includegraphics[width=0.82\linewidth]{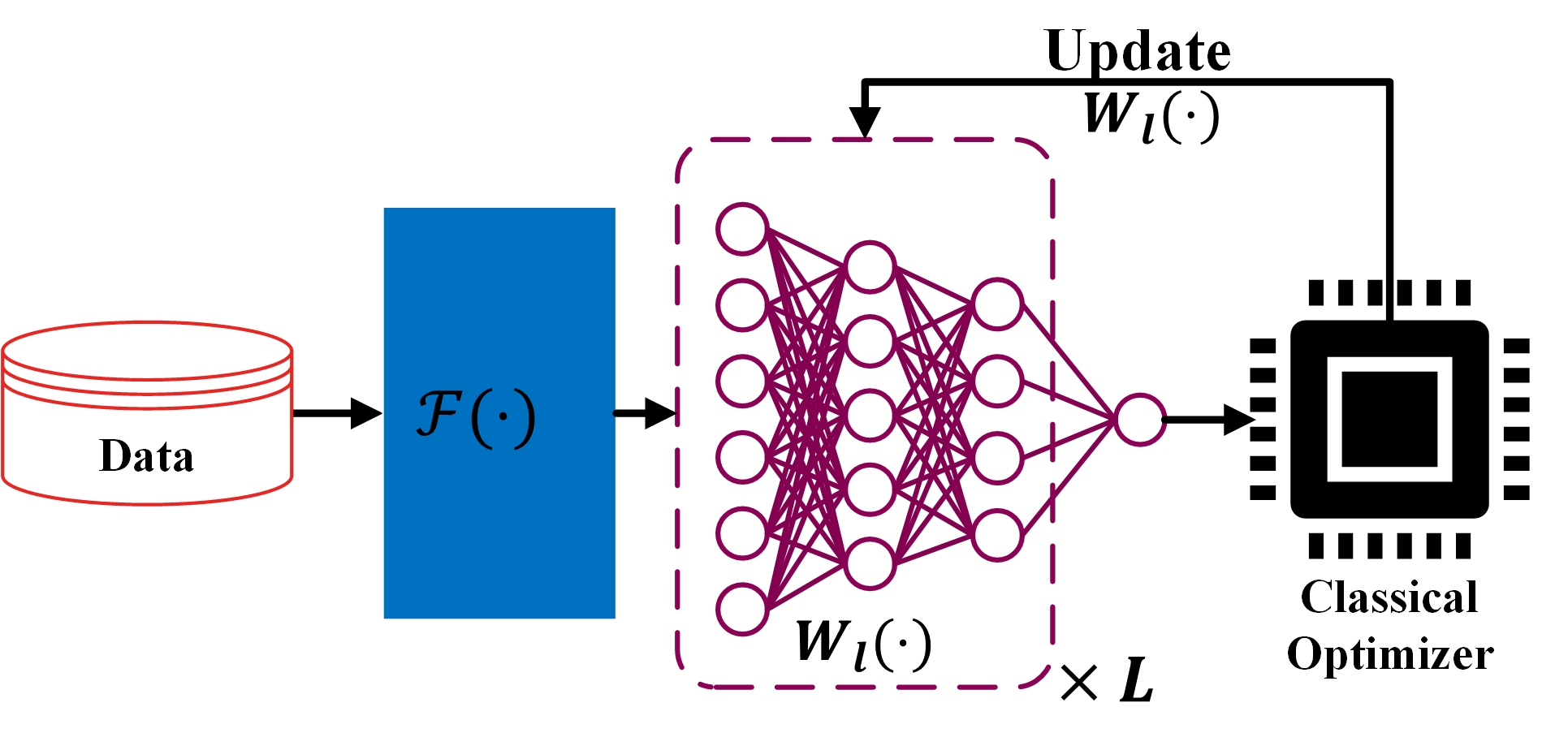}
	\vspace{-0.3cm}
	\caption{Classical neural network model}
	\label{fig: Classical_ANN}
\end{figure}

Figure \ref{fig: Classical_ANN} presents a typical neural network used for regression in classical computers. 
The dataset, e.g., the empirical data of battery, can be first fed to the feature embedded layer $\mathcal{F}(.)$, which consists of a sequence of operations \& combination (e.g., convolution and attention) to be mapped to the feature space.
Then the hidden layer $W_l(.)$, typically $l$-fully connected,  takes featured data and produces the estimation results.  
The regression problem boils down to the problem of tuning weight factors of neurons in $W_l(.)$ such that the loss function (e.g., RMSE or MAPE) is minimized.
This can be addressed by several effective numerical algorithms developed in the literature such as  BFGS (Broyden–Fletcher–Goldfarb–Shannon algorithm), ADAM (Adaptive Movement Estimation algorithm), and Nelder-Mead \cite{goodfellow2016deep}.



\begin{figure*}[t!]
\centering
\subfigure[The Bloch sphere]{\includegraphics[width=0.25\linewidth]{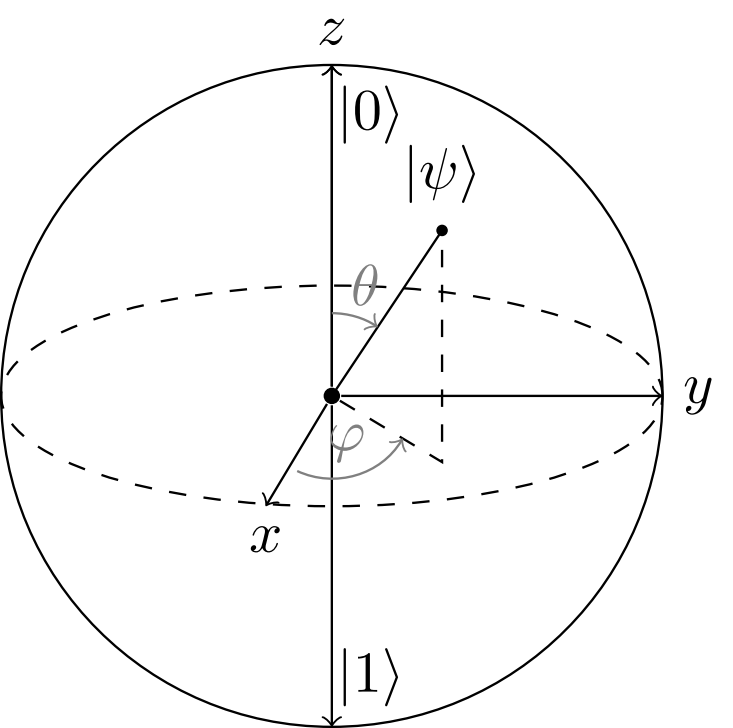} \label{fig: The Bloch sphere}} 
%
\subfigure[QNN Work Flow]{\includegraphics[width=0.73\linewidth]{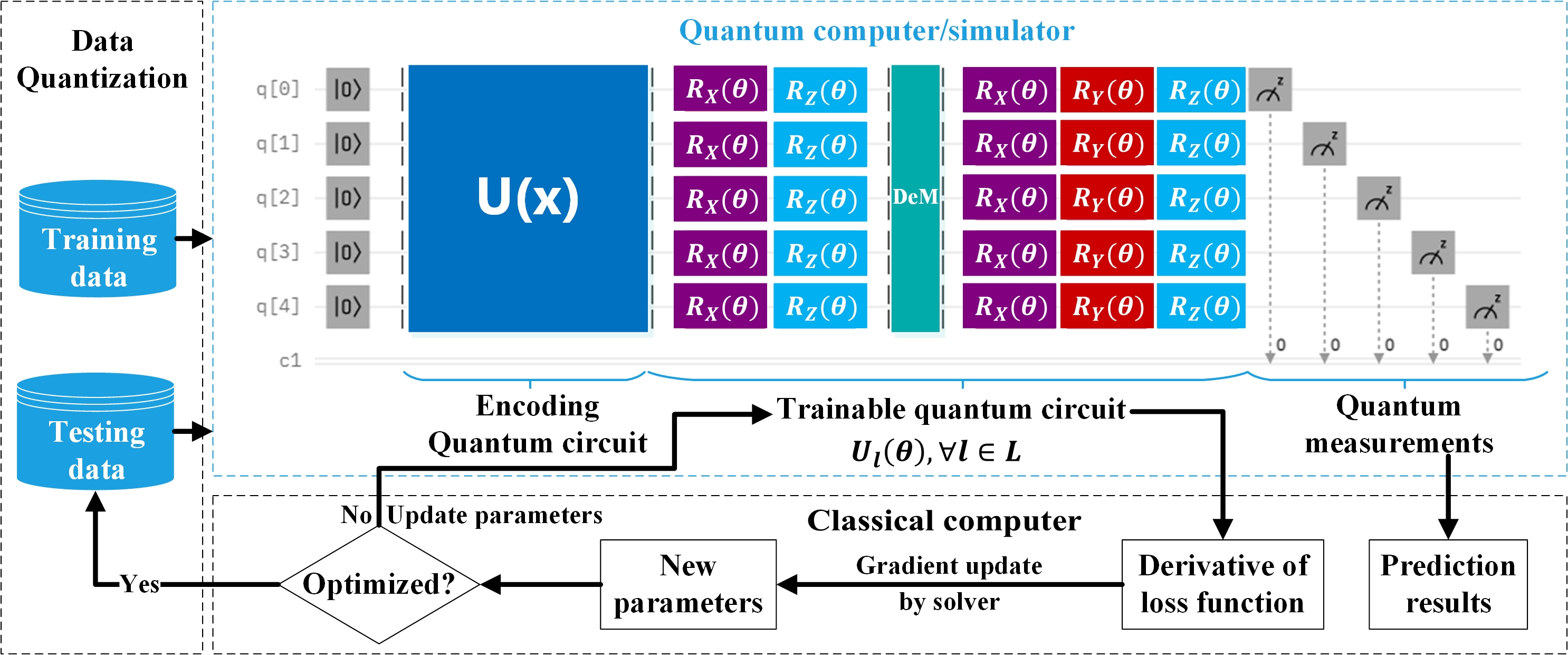}
\label{fig: QNN}
}%
\caption{Estimate Lithium-ion battery degradation using Quantum Neural Network}
\label{fig: QNN_System_Model}
\end{figure*}

\section{Quantum neural network regression}



\subsection{Overview of Quantum Neural Network}
Quantum neural networks are motivated by (i) the huge success of neural networks, particularly deep learning, in solving a real-world problem and (ii) the recent advances in quantum computing \cite{quantum_circuit_learning,qcl2}.
It emerges following the similarity between the quantum circuit and the neural network architecture as illustrated in Figure \ref{fig: QNN_System_Model}.

\subsubsection{Quantum bit}
QNN utilizes a quantum bit (qubit) as a neuron in the classical neural network. 
Unlike a binary digit in classical computers,
a single qubit can be in a superposition of two basis states $|0\rangle$ and $|1\rangle$ \cite{quantum_info}, i.e., it can be represented in a two-dimensional complex vector space as follows:
\beqn
| \psi \rangle = \alpha | 0 \rangle + \beta | 1 \rangle : \alpha, \beta \in \mathbb{C} \wedge |\alpha|^2 + |\beta|^2 = 1.
\eeqn
Consequently, the state of a qubit can be ininventionted as a point on a complex sphere with radius 1, i.e., the Bloch sphere as in Figure \ref{fig: The Bloch sphere}.
In other words, its state can be represented by two angles $\theta$ and $\varphi$, which are called two degrees of freedom, and tuning $\theta$ can alternate the state of the qubit.

\subsubsection{Quantum Circuit}
A quantum circuit is a simple sequence of quantum gates, measurements, initialization of qubits to known values, and other actions which are represented to a quantum computation. 
On a circuit, a sequence of qubits is encoded as $q_0,q_1,\dots,q_{n-1}$, in which $n$ is the number of qubits.
The qubit can be a result rotated along $x-y-z$ axes for the probability distributions of  $| 0 \rangle$ and $ | 1 \rangle $. 
The probability depends on the internal state of the qubit, but more specifically, on the angle $\theta$ between state vector $ | \psi\rangle $ and the measured axis.
Therefore, a numerical feature can be approximated into a qubit by $f: \mathrm{R} \rightarrow [0, \pi]$.
The states of the qubit can be modulated by the quantum gates, such as a single qubit gate (e.g, Pauli-X gate) or multiple-qubit gate (e.g, two-qubit CNOT gate). 
Quantum gates are unitary operators and are illustrated as unitary matrices represented on some computational basis.
Here, $R_X$, $R_Y$, and $R_Z$ gates are rotation operators representing a single-qubit rotation through angle $\theta$ (radians) around the $x$-axis, $y$-axis, and $z$-axis, respectively:
\begin{align}
R_X (\theta) &= 
\begin{bmatrix}
    \cos{\frac{\theta}{2}} & -i\sin{\frac{\theta}{2}} \\
    \-i\sin{\frac{\theta}{2}} & cos{\frac{\theta}{2}}
\end{bmatrix}
\\
R_Y (\theta) &= 
\begin{bmatrix}
    \cos{\frac{\theta}{2}} & -\sin{\frac{\theta}{2}} \\
    \\sin{\frac{\theta}{2}} & cos{\frac{\theta}{2}}
\end{bmatrix}
\\
R_Z (\theta) &= 
\begin{bmatrix}
    e^{-i\frac{\theta}{2}} & 0\\
    0 & e^{i\frac{\theta}{2}}
\end{bmatrix}
\end{align}
A digital error model (DeM) is used to control the quantum gates, and readout errors and characterizes circuit performance by a set of Pauli errors \cite{quantum_kernel_ibm2022}.
Finally, a measurement component is compulsory to read out the result of the quantum computation. 
The output of the measurement component is a classical variable value retrieved from a probability distribution governed by the Born rule of the quantum theory \cite{quantum_info}.





\subsection{QNN Implementation and training procedure}
The structure of QNN represented in Figure \ref{fig: QNN} follows the typical structure of the neural network shown in Figure \ref{fig: Classical_ANN}. 
The first stage, which resembles the feature-embedded layer in the classical neural network, is \textit{the encoding quantum circuit} $U(X)$.
It is used to encode classical data to be used in the quantum computer.
This stage plays a vital role in adopting quantum algorithms to solve classical problems, particularly in quantum deep learning tasks.
\textit{The trainable quantum circuit} resembles the classical counterpart's hidden layer where tuning circuit parameters $\theta$ results in the change of the output in the quantum measurement.
\textit{The quantum measurement}, in turn, acts like the output layer in the classical neural network, i.e., its output after decoding represents the prediction result.

The quantum circuit of the QNN model can be run in a quantum simulator or actual quantum hardware (e.g., IBM Quantum servers, Amazon Braket, Rigetti Computing, D-Wave, and Strawberry Fields) \cite{Q_annealing_Powersyst, quan_noise_Pow_Syst}. 
However, the availability of quantum hardware is currently limited, e.g., the IBM server only allows 7 qubits for free application  \cite{NAPS2022, qiskit}. 
The computer simulator back-end for quantum circuit become useful for proof-of-concept research.  
This paper uses the quantum simulator back-end provided by Qulacs \cite{qulacs} and trains the QNN using the hybrid classical-quantum procedure with BFGS optimizer \cite{quantum_circuit_learning}:
\begin{itemize}
    \item Prepare training data and encode the training data into the quantum state vectors $|\psi_{in} \rangle$ using angle encoding. The input state is obtained as $|\psi_{in} \rangle = U(x)$.
    \item $\theta$-parameterized unitary $U(\theta)$ is applied to the input state and generates an output state $|\psi_{out} \rangle$.
    \item Measure the expectation values of some chosen observable. For instance, $Z$ expectation of the second qubit is denoted as $\langle Z_2 \rangle = \langle \psi_{out} | Z_2 | \psi_{out} \rangle$.
    \item Minimize the cost function $L$ of between the real measured capacity $y_i$ and the prediction $\hat{y}_i$ by tuning the quantum circuit parameter $\theta$ iterative until $\theta = \theta^*$. In which, $y(x,\theta^*)$ is the desired prediction model.
    \item Evaluate the accuracy of the QNN regression model by validating the cost function with a testing data set.
\end{itemize}

\subsection{Potential advantages of the quantum neural  network over classical neural network}
Although quantum neural networks (QNN) and their classical counterparts have similar features, there are some differences, resulting in the possible QNN advantages \cite{q_neural_design}.  
In order to construct a complex model for high-precision prediction, classical neural networks require the use of nonlinear basis functions or the use of kernel tricks,
leading to a significant increase in computational cost. 
In contrast, leveraging quantum mechanics, QNN can directly employ the exponential number of functions with respect to the number of qubits, thus capturing non-linearity without using nonlinear activation functions. 
Indeed, the trainable quantum circuit $U(\theta)$ already has strong expressive power, which potentially enables a wide range of applications in capturing nonlinearity that are intractable on classical counterparts \cite{VQA_classifier}. 
%



\section{Numerical results}
\subsection{Li-ion battery data pre-processing}
The datasets used in this research are collected from the NASA Prognostics Center of Excellence Data Set Repository \cite{nasa_battery}.
There are 34 different batteries in the datasets that comprise more than 6  profiles in charging and discharging states conducted on Lithium Nickel Manganese Cobalt Oxide batteries. 
Batteries were charged and discharged at different temperatures, and the impedance was measured at every cycle.
The batteries have 18650 Lithium-ion cells with 2-Ah capacity for B05, B06 and B18, and 1.35-Ah capacity for B56, for which the charging and discharging cycle experiments were conducted repeatedly to achieve accelerated aging.
A battery with state of health (SoH)  reducing below $70 \%$ is considered to be discarded because its operational performance is not reliable \cite{Classical_Battery_ML}.
In other words, the experiments are stopped when the battery loses over $30 \%$ of its rated capacity.

\begin{figure}[http]
    \centering
    \subfigure{\includegraphics[width=0.82\linewidth]{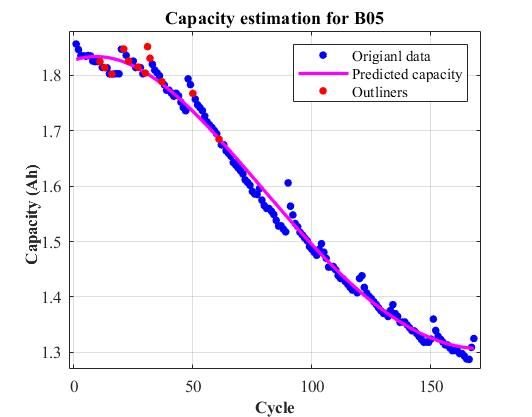} \label{fig: SingPred05}}  
    \vspace{-0.25cm}
    \\
        \subfigure{\includegraphics[width=0.82\linewidth]{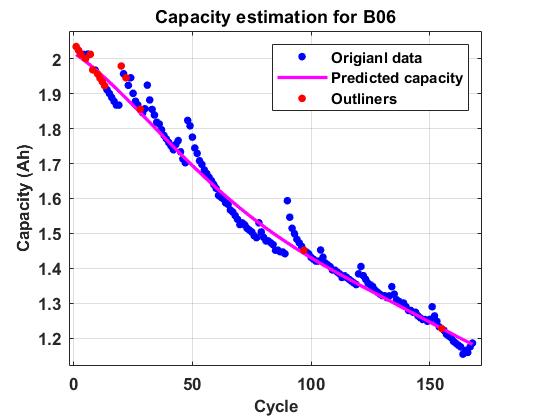} \label{fig: SingPred06}} 
        \vspace{-0.25cm}
    \\
        \subfigure{\includegraphics[width=0.82\linewidth]{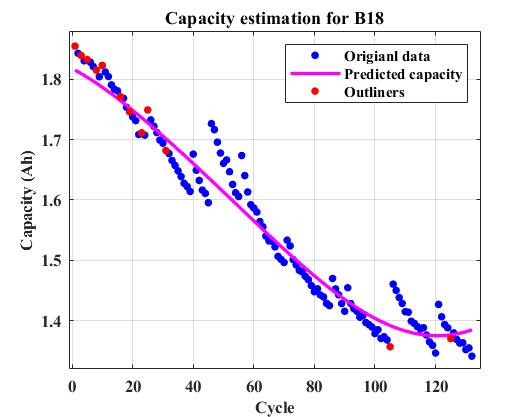} \label{fig: SingPred18}} 
        \vspace{-0.25cm}
    \\
        \subfigure{\includegraphics[width=0.82\linewidth]{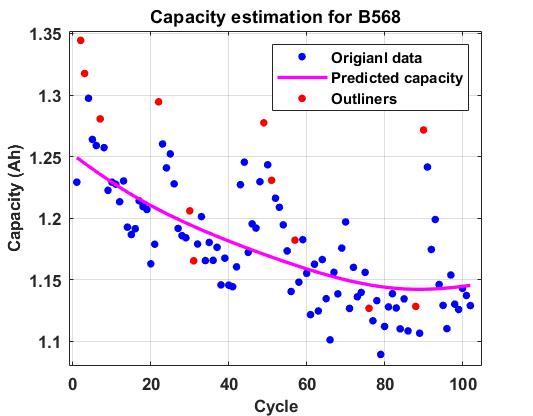} \label{fig: SingPred56}} 
	\vspace{-0.3cm}
	\caption{Estimation of Lithium-ion batteries' degradation using QNN}
	\label{fig: SingPred}
\end{figure}

\subsection{QNN Regression Model}
We chose the batteries B05, B06, B18, and B56 to validate the quantum regression model and the quantum circuit.
Here, data of B05, B06, and B18 are quite clustered while data of B56 has a lot of noise.
The train size was chosen at 80\% of the original data for the first experiments on B05, B06, B18, and B56.
We conduct the experiments on the quantum simulator powered by a classical computer with a configuration of 32 GB RAM and an Intel Xeon processor.
Capacity aging prediction patterns and the original capacity measurements from the four LiBs demonstrated in Figure \ref{fig: SingPred} are positively correlated.
The QNN model works well with various data with different characteristics, neither over-fitted nor under-fitted with good RMSE and MAPE metrics

\begin{figure}[t!]
    \centering
	\includegraphics[width=0.82\linewidth]{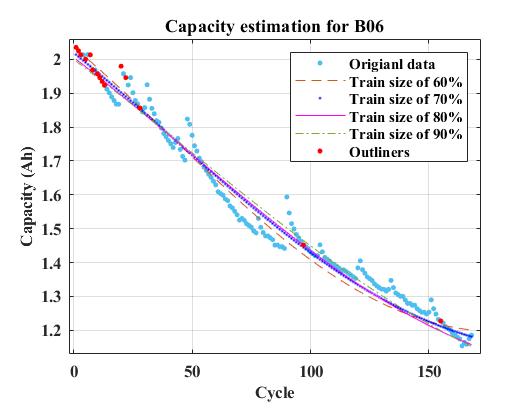}
	\vspace{-0.4cm}
	\caption{Estimation of Lithium-ion battery degradation using QNN with different training sizes}
	\label{fig: MuPred_B06}
\end{figure}

We examine the impacts of training data size on the performance of QNN.
Figure \ref{fig: MuPred_B06} shows results for cross-validation of the accuracy of capacity degradation estimation based on the QNN for several different sizes (i.e., 60\%, 70\%, 80\% and 90\%) of the training sample.
The obtained RMSE and MAPE are quite small, which highlights the ability of the neural network empowered by the quantum circuit in capturing the nonlinear relationship of battery capacity on the operational cycle. 
As different kinds of capacity degradation paths are shown in the figure, we can see that the larger data size also influences on the prediction results.
In detail, the larger size of the training data set leads to a more coherent prediction.

\begin{figure}[t!]
    \centering
	\includegraphics[width=0.8\linewidth]{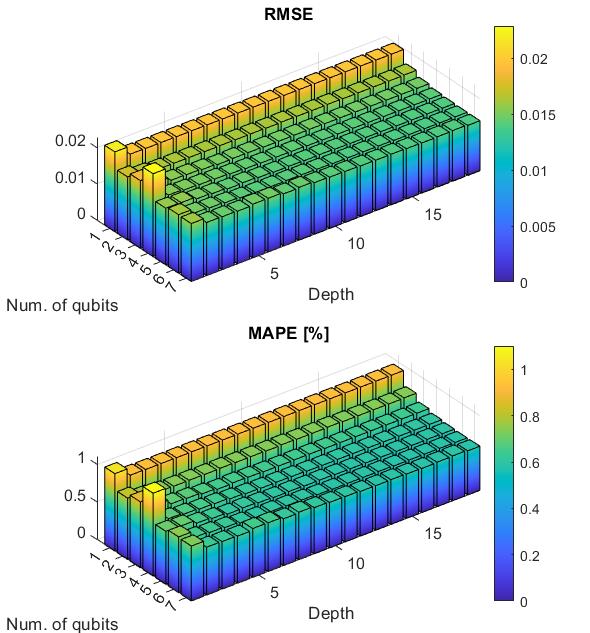}
	\vspace{-0.2pt}
	\caption{Effect of the number of qubits and the depth on the accuracy of the QNN model on B06}
	\label{fig: q_and_depth_b07}
\end{figure}

Figure \ref{fig: q_and_depth_b07} shows the correlation between the depth and the number of qubits with the accuracy of the QNN model.
In particular, the depth and the number of qubits are considered as hypo-parameters of QNN since they control the architecture or topology of the QNN.
The obtained results show that the qubits and depth quantities increase to a certain amount, and the RMSE and MAPE indicators drop insignificantly.
Meanwhile, a quantum circuit with a large number of qubits and depth is computationally expensive.
In other words, similar to the classical neural network, there is a trade-off in the complexity of the QNN and the obtained performance, which should be considered carefully in the design phase. 

\section{Conclusion}
This paper studied the quantum neural network regression model for the predicting battery life in which the qubit can resemble the neuron in classical machine learning. 
The QNN inherits the principle and structure of the classical neural network and the training of QNN boils into tuning the parameters of qubits in the quantum circuits, which is similar to tuning weights of classical neurons. 
The QNN's expression ability can be empowered thanks to the capability of a qubit in capturing complex nonlinearity.  
Similar to the classical neural network, the performance of QNN also depends on design parameters such as the number of qubits and depth of the quantum circuit.
Numerical results show that QNN can be a very promising approach for modeling battery usage, especially for the future penetration of EVs and energy storage under the decarbonization policy.

\bibliographystyle{IEEEtran}
\bibliography{tpec2023_quantum}

\begin{thebibliography}{10}
\providecommand{\url}[1]{#1}
\csname url@samestyle\endcsname
\providecommand{\newblock}{\relax}
\providecommand{\bibinfo}[2]{#2}
\providecommand{\BIBentrySTDinterwordspacing}{\spaceskip=0pt\relax}
\providecommand{\BIBentryALTinterwordstretchfactor}{4}
\providecommand{\BIBentryALTinterwordspacing}{\spaceskip=\fontdimen2\font plus
\BIBentryALTinterwordstretchfactor\fontdimen3\font minus
  \fontdimen4\font\relax}
\providecommand{\BIBforeignlanguage}[2]{{%
\expandafter\ifx\csname l@#1\endcsname\relax
\typeout{** WARNING: IEEEtran.bst: No hyphenation pattern has been}%
\typeout{** loaded for the language `#1'. Using the pattern for}%
\typeout{** the default language instead.}%
\else
\language=\csname l@#1\endcsname
\fi
#2}}
\providecommand{\BIBdecl}{\relax}
\BIBdecl

\bibitem{improved_softplus_ANN}
J.~Wang, X.~Feng, X.~Zhang, and Y.~Xiang, ``Improved modeling of lithium-ion
  battery capacity degradation using an individual-state training method and
  recurrent softplus neural network,'' \emph{IEEE Access}, vol.~9, pp.
  7845--7855, 2020.

\bibitem{bat_cap_degradation2021}
F.~Zhou and C.~Bao, ``Analysis of the lithium-ion battery capacity degradation
  behavior with a comprehensive mathematical model,'' \emph{Journal of Power
  Sources}, vol. 515, p. 230630, 2021.

\bibitem{laresgoiti2015modeling}
I.~Laresgoiti, S.~K{\"a}bitz, M.~Ecker, and D.~U. Sauer, ``Modeling mechanical
  degradation in lithium ion batteries during cycling: Solid electrolyte
  interphase fracture,'' \emph{Journal of Power Sources}, vol. 300, pp.
  112--122, 2015.

\bibitem{review_on_ML_methods}
X.~Cong, C.~Zhang, J.~Jiang, W.~Zhang, and Y.~Jiang, ``A hybrid method for the
  prediction of the remaining useful life of lithium-ion batteries with
  accelerated capacity degradation,'' \emph{IEEE Transactions on Vehicular
  Technology}, vol.~69, no.~11, pp. 12\,775--12\,785, 2020.

\bibitem{Classical_Battery_ML}
Y.~Choi, S.~Ryu, K.~Park, and H.~Kim, ``Machine learning-based lithium-ion
  battery capacity estimation exploiting multi-channel charging profiles,''
  \emph{Ieee Access}, vol.~7, pp. 75\,143--75\,152, 2019.

\bibitem{goodfellow2016deep}
I.~Goodfellow, Y.~Bengio, and A.~Courville, \emph{Deep learning}.\hskip 1em
  plus 0.5em minus 0.4em\relax MIT press, 2016.

\bibitem{hybrid_qnn}
P.~Reddy and A.~B. Bhattacherjee, ``A hybrid quantum regression model for the
  prediction of molecular atomization energies,'' \emph{Machine Learning:
  Science and Technology}, vol.~2, no.~2, p. 025019, 2021.

\bibitem{quantum_kernel_ibm2022}
Z.~Krunic, F.~F. Fl{\"o}ther, G.~Seegan, N.~D. Earnest-Noble, and O.~Shehab,
  ``Quantum kernels for real-world predictions based on electronic health
  records,'' \emph{IEEE Transaction on Quantum Engineering}, vol.~3, pp. 1--11,
  2022.

\bibitem{quantum_info}
M.~A. Nielsen and I.~Chuang, ``Quantum computation and quantum information,''
  2002.

\bibitem{q_neural_design}
Y.~Zhang and Q.~Ni, ``Design of quantum neuron model for quantum neural
  networks,'' \emph{Quantum Engineering}, vol.~3, no.~3, p. e75, 2021.

\bibitem{quantum_circuit_learning}
K.~Mitarai, M.~Negoro, M.~Kitagawa, and K.~Fujii, ``Quantum circuit learning,''
  \emph{Physical Review A}, vol.~98, no.~3, p. 032309, 2018.

\bibitem{qcl2}
M.~Schuld, V.~Bergholm, C.~Gogolin, J.~Izaac, and N.~Killoran, ``Evaluating
  analytic gradients on quantum hardware,'' \emph{Physical Review A}, vol.~99,
  no.~3, p. 032331, 2019.

\bibitem{reason4bat_cap_degradation2020}
X.~Li, X.~Sun, X.~Hu, F.~Fan, S.~Cai, C.~Zheng, and G.~D. Stucky, ``Review on
  comprehending and enhancing the initial coulombic efficiency of anode
  materials in lithium-ion/sodium-ion batteries,'' \emph{Nano Energy}, vol.~77,
  p. 105143, 2020.

\bibitem{Q_annealing_Powersyst}
T.~Krauss, J.~McCollum, C.~Pendery, S.~Litwin, and A.~J. Michaels, ``Solving
  the max-flow problem on a quantum annealing computer,'' \emph{IEEE
  Transactions on Quantum Engineering}, vol.~1, pp. 1--10, 2020.

\bibitem{quan_noise_Pow_Syst}
Y.~Zhou and P.~Zhang, ``Noise-resilient quantum machine learning for stability
  assessment of power systems,'' \emph{IEEE Transactions on Power Systems},
  2022.

\bibitem{NAPS2022}
P.~Ngo, C.~Thomas, H.~Nguyen, A.~Eroglu, and K.~Oikonomou, ``Evaluate quantum
  combinatorial optimization for distribution network reconfiguration,''
  \emph{North American Power Symposium}, 2022.

\bibitem{qiskit}
\BIBentryALTinterwordspacing
{IBM Qiskit Development Team}, ``Qiskit: An open-source framework for quantum
  computing,'' 2022. [Online]. Available:
  \url{https://qiskit.org/documentation}
\BIBentrySTDinterwordspacing

\bibitem{qulacs}
Y.~Suzuki, Y.~Kawase, Y.~Masumura, Y.~Hiraga, M.~Nakadai, J.~Chen, K.~M.
  Nakanishi, K.~Mitarai, R.~Imai, S.~Tamiya \emph{et~al.}, ``Qulacs: a fast and
  versatile quantum circuit simulator for research purpose,'' \emph{Quantum},
  vol.~5, p. 559, 2021.

\bibitem{VQA_classifier}
H.~Yano, Y.~Suzuki, R.~Raymond, and N.~Yamamoto, ``Efficient discrete feature
  encoding for variational quantum classifier,'' in \emph{2020 IEEE
  International Conference on Quantum Computing and Engineering (QCE)}.\hskip
  1em plus 0.5em minus 0.4em\relax IEEE, 2020, pp. 11--21.

\bibitem{nasa_battery}
B.~Saha and K.~Goebel, ``Batter data set,'' in \emph{Prognostics Center of
  Excellence Data Set Repository}, no.~5.\hskip 1em plus 0.5em minus
  0.4em\relax NASA Ames Research Center, Moffett Field, CA, 2003.

\end{thebibliography}

\end{document}